\documentclass[twocolumn]{aastex61}
\usepackage{amsmath}

\newcommand{\bdv}[1]{\mbox{\boldmath$#1$}}

\def\au{{\rm AU}}

\def\mas{{\rm mas}}
\def\sat{{\rm sat}}
\def\muas{\mu{\rm as}}

\def\obs{{\rm obs}}

\def\min{{\rm min}}
\def\rel{{\rm rel}}

\def\eff{{\rm eff}}

\def\e{{\rm E}}
\def\bpi{{\bdv\pi}}
\def\bmu{{\bdv\mu}}

\def\btheta{{\bdv\theta}}

\def\bv{{\bf v}}
\def\bu{{\bf u}}

\begin{document}
\title{First Resolution of Microlensed Images\footnote{Based on observations made with ESO telescopes at Paranal 
observatory under program ID 2100.C-5014.}}

\correspondingauthor{Subo Dong}

\email{dongsubo@pku.edu.cn}

\author{Subo Dong}\affil{Kavli Institute for Astronomy and Astrophysics, Peking University, Yi He Yuan Road 5, Hai Dian District, Beijing 100871, China}
\author{A.~M\'erand}\affil{European Southern Observatory, Karl-Schwarzschild-Str. 2, 85748 Garching, Germany}
\author{F.~Delplancke-Str\"obele}\affil{European Southern Observatory, Karl-Schwarzschild-Str. 2, 85748 Garching, Germany}
\author{Andrew Gould}\affil{Max-Planck-Institute for Astronomy, K\"onigstuhl 17, 69117 Heidelberg, Germany}\affil{Korea Astronomy and Space Science Institute, Daejon 305-348, Republic of Korea}\affil{Department of Astronomy Ohio State University, 140 W.\ 18th Ave., Columbus, OH 43210, USA}
\author{Ping Chen}\affil{Kavli Institute for Astronomy and Astrophysics, Peking University, Yi He Yuan Road 5, Hai Dian District, Beijing 100871, China}
\author{R. Post}\affil{Post Observatory, Lexington, MA 02421}
\author{C.~S. Kochanek}\affil{Department of Astronomy Ohio State University, 140 W.\ 18th Ave., Columbus, OH 43210, USA} \affil{Center for Cosmology and AstroParticle Physics (CCAPP), The Ohio State University, 191 W. Woodruff Avenue, Columbus, OH 43210, USA.} 
\author{K.~Z. Stanek}\affil{Department of Astronomy Ohio State University, 140 W.\ 18th Ave., Columbus, OH 43210, USA} \affil{Center for Cosmology and AstroParticle Physics (CCAPP), The Ohio State University, 191 W. Woodruff Avenue, Columbus, OH 43210, USA.}
\author{G.~W. Christie}\affil{Auckland Observatory, Box 24180, Auckland, New Zealand}
\author{Robert Mutel}\affil{Department of Physics and Astronomy, University of Iowa}
\author{T. Natusch}\affil{Institute for Radio Astronomy and Space Research, AUT University, Auckland, New Zealand}
\author{T.~W.-S. Holoien}\affil{The Observatories of the Carnegie Institution for Science, 813 Santa Barbara St., Pasadena, CA 91101, USA 0000-0001-9206-3460}
\author{J.~L. Prieto}\affil{N\'ucleo de Astronom\'ia de la Facultad de Ingenier\'ia y Ciencias, Universidad Diego Portales, Av. Ej\'ercito 441, Santiago, Chile}\affil{Millennium Institute of Astrophysics, Santiago, Chile}
\author{B.~J. Shappee}\affil{Institute for Astronomy, University of Hawai'i, 2680 Woodlawn Drive, Honolulu, HI 96822, USA}
\author{Todd~A. Thompson}\affil{Department of Astronomy Ohio State University, 140 W.\ 18th Ave., Columbus, OH 43210, USA} \affil{Center for Cosmology and AstroParticle Physics (CCAPP), The Ohio State University, 191 W. Woodruff Avenue, Columbus, OH 43210, USA.}

\begin{abstract}

We employ VLTI GRAVITY to resolve, for the first time,
the two images generated by a gravitational microlens.
The measurements 
of the image separation $\Delta\theta_{-,+} =3.78\pm 0.05\,$mas, 
and hence the Einstein radius $\theta_{\rm E}=1.87\pm 0.03\,$mas, 
are precise.  This demonstrates the robustness of the method,
provided that the source is bright enough for GRAVITY $(K \lesssim 10.5)$
and the image separation is of order or larger than the fringe
spacing.  When $\theta_\e$ is combined with a measurement of the
``microlens parallax'' $\pi_\e$, the two will together yield the
lens mass and lens-source relative parallax and proper motion.  Because
the source parallax and proper motion are well measured by Gaia, this
means that the lens characteristics will be fully determined, whether
or not it proves to be luminous.  This method can be a powerful
probe of dark, isolated objects, which are otherwise quite difficult
to identify, much less characterize.  Our measurement contradicts
Einstein's (1936) prediction that ``the luminous circle [i.e., microlensed
image] cannot be distinguished'' from a star.

\end{abstract}


\section{{Introduction}
\label{sec:intro}}

Interferometric resolution of microlensed images is a potentially powerful probe of dark isolated objects \citep{vltimicrolens1}.  Orbiting dark objects, such as planets, black holes, and old brown dwarfs and neutron stars, can be studied by a variety of techniques, via their impact on their companions. For example, all of these objects can be studied through their gravitational effect on their hosts by radial-velocity and astrometric methods.  Planets and brown dwarfs can be studied  via their occulting effects using the transit method.  
Some black holes and neutron stars in binaries can be discovered and characterized because they are accreting gas from stellar companions. Even double-dark objects like black-hole and neutron-star binaries can be detected and studied with gravitational waves.

However, the only known way to study isolated dark objects is with gravitational microlensing.  In microlensing, a massive object temporarily magnifies the light of a more distant source in a ``microlensing event''.  The object can therefore be detected independent of whether it is dark or luminous.  The key challenge of microlensing is that the only parameter that can be routinely measured in microlensing events is the Einstein timescale $t_\e$, which is a combination of the lens mass, $M$, and the lens-source relative (parallax, proper motion), $(\pi_\rel,\bmu_\rel)$.
\begin{equation}
t_\e={\theta_\e\over\mu_\rel};
\qquad
\theta_\e\equiv\sqrt{\kappa M\pi_\rel};
\qquad
\kappa\equiv {4 G\over c^2\au}\simeq 8.14{\mas\over M_\odot};
\label{eqn:tedef}
\end{equation}
where $\theta_\e$ is called the angular Einstein radius, which is on the order of mas for Galactic microlensing.
Hence, for dark microlenses, the only way to recover these three quantities, $(M,\pi_\rel,\bmu_\rel)$, separately is to measure $\theta_\e$ and the ``microlens parallax'' $\bpi_\e$ \citep{refsdal66,gould92,gould00,gould04},
\begin{equation}
\bpi_\e \equiv {\pi_\rel\over\theta_\e}{\bmu_\rel\over\mu_\rel} .
\label{eqn:bpie}
\end{equation}
By measuring these parameters ($\theta_\e$ and $\bpi_\e$),
one can then determine
\begin{equation}
M={\theta_\e\over\kappa\pi_\e};
\qquad
\pi_\rel = \theta_\e\pi_\e;
\qquad
\bmu_\rel = {\theta_\e\over t_\e}\,{\bpi_\e\over\pi_\e}.
\label{eqn:mpirelmurel}
\end{equation}
Then, if the source parallax and proper motion $(\pi_s,\bmu_s)$ can also be measured, one can determine the lens distance and transverse velocity as well.

For dark lenses, there are only three known ways to measure $\theta_\e$, and each poses significant challenges.  Prior to this paper, the method that had been most successfully applied was to measure the so-called 
``finite-source effects'' as the lens transits the source  \citep{gould94a}.  This yields the microlensing parameter 
$\rho=\theta_*/\theta_\e$, where $\theta_*$ is the angular radius of the source, which can usually be estimated quite well from its color and magnitude \citep{ob03262}.  The main problem with this method is that the probability that the lens will transit the source for
any given microlensing event is equal to $\rho$, which typically has
values $\rho\sim 10^{-2}$--$10^{-3}$.

A second method to measure $\theta_\e$ is ``astrometric microlensing'',
wherein one measures the astrometric displacement of the light centroid
of the microlensed images from the position of the source. 
Isolated objects create two images whose offsets from the source
and magnifications are given by \citep{einstein36,pac86}
\begin{equation}
\btheta_\pm = u_\pm\theta_\e \,{\bu\over u};
\qquad
u_\pm \equiv {u\pm\sqrt{u^2+4}\over 2},
\label{eqn:upm}
\end{equation}
and
\begin{equation}
A_\pm = {A\pm 1\over 2};
\qquad
A = {u^2+2 \over u\sqrt{u^2 +4}},
\label{eqn:apm}
\end{equation}
where $\bu$ is the vector offset of the source from the lens, normalized
by $\theta_\e$.
Hence, the centroid of the light from the images is offset from the position of the source by
\citep{my95,hnp95,walker95}.
\begin{equation}
\Delta\btheta_{i,s} = \biggl({A_+\theta_+ + A_-\theta_-\over A} - u\biggr)
{\bu\over u} = {{\bu}\over u^2 + 2}\theta_\e.
\label{eqn:astromic}
\end{equation}
This method of course requires a time series of astrometric measurements
because one must simultaneously solve for $(\pi_s,\bmu_s)$ in order
to determine the intrinsic source position as a function of time, which
is a precondition for measuring the ``offset'' from that position.
However, this is also an advantage because, as mentioned above,
these quantities are needed to infer the lens distance and transverse
velocity once $(\theta_\e,\bpi_\e)$ are measured.

Another advantage of this technique, as discussed by \citet{gould14},
is that it actually measures two of the three parameters $(\theta_\e,\bpi_\e)$
needed for the mass measurement.  That is, it measures both $\theta_\e$
and the {\it direction} of $\bpi_\e$.  The reason that this is important
is that of the two components of $\bpi_\e$, the component that is parallel
to Earth's acceleration is much easier to measure than the component that
is perpendicular \citep{gmb94,smp03,gould04}.  Hence, if the direction
of $\bpi_\e$ can be independently determined, this dramatically increases
the prospects for measuring $\bpi_\e$ \citep{ghosh04, gould15}.  While there has been 
only two reported successful application to date using the 
{\it Hubble Space Telescope} (HST) by \citet{sahu17} and with HST and VLT/Spectro-Polarimetric High-contrast Exoplanet REsearch (SPHERE) by \citet{sphere}, the prospects for 
making such measurement with {\it Gaia} \citep{gaia} are very good as we discuss below (also see, e.g., \citealt{gaiaastr,gaiaastr2}).

Here we present the first application of a third method to determine 
$\theta_\e$, interferometric resolution of the microlensed images. 
{As anticipated by \citet{einstein36}, the mas scale of $\theta_E$ for a Galactic microlensing event is far smaller than the resolution of any existing or planned optical telescope for direct resolution of the images.  It is, however, possible to do so using interferometry, combining the light from well-separated telescopes to give a much higher angular resolution than the individual elements.  
Interferometry is routine in radio astronomy and well-developed for 
small optical telescopes, but has only recently started to become 
available on large optical telescopes.  It is still very challenging and 
requires a very rare, bright microlensing event.   
The handful of existing publications on this method \citep{vltimicrolens1,vltimicrolens2,vltimicrolens3,vltimicrolens4} mainly focus on forecasting its prospects, while here we describe how the actual observational data can be analyzed to constrain microlensing parameters.}
Like the
``astrometric microlensing'' method just discussed, it has the important
advantage that it can simultaneously measure the direction of $\bpi_\e$.
Stated more succinctly, it can measure $\bmu_\rel$.

The basic idea of the measurement is straightforward: simply measure
the vector separation between the two images, i.e., from the minor image
to the major image, $\Delta\btheta_{-,+}$.  The magnitude of this
separation,
\begin{equation}
\Delta\theta_{-,+} = \sqrt{u^2+4}\theta_\e = 
2\biggl(1 + {u^2\over 8} \ldots\biggr)\theta_\e ,
\label{eqn:dthetamp}
\end{equation}
directly yields $\theta_\e$, given that $u$ is known from the photometric
light curve (note that the interferometric data also allow to measure $u$ 
from the flux ratio between the two lensed images).  In fact, as can be seen from the Taylor expansion in
Equation~(\ref{eqn:dthetamp}), for cases that the minor image
can be detected with reasonable effort ($u\la 1/2$, see 
Equation~(\ref{eqn:apm})), even very crude knowledge of $u$ is sufficient
for an accurate measurement of $\theta_\e$.

The direction of $\bpi_\e$ must be derived from the measurement of
$\Delta\btheta_{-,+}$ in two steps.  In itself, $\Delta\btheta_{-,+}$
only gives the direction of the instantaneous separation $\bu$.
Then, one must combine this with the angle between $\bu$ and $\bmu_\rel$
to obtain the direction of source motion (or of $\bpi_\e$).  This
angle $\phi$ is given by
\begin{equation}
\phi(u) = \cot^{-1}{(t_\obs - t_0)/t_\e \over u_0}
= \cot^{-1}{\delta t \over t_\eff},
\label{eqn:phiu}
\end{equation}
and $\phi$ is always defined to be positive and between 0 and $\pi$.
Here, $(t_0,u_0,t_\e)$ are the \citet{pac86} parameters describing the
trajectory, i.e., $u^2(t) = u_0^2 + (t-t_0)^2/t_\e^2$, $t_\obs$ is the
time of the observation, $\delta t\equiv t_\obs - t_0$, and 
$t_\eff = u_0 t_\e$ is the effective timescale.

The final form of Equation~(\ref{eqn:phiu}) is very important.
In general, $t_0$ is much better determined than either $u_0$ or
$t_\e$ because the latter two are strongly anti-correlated and are
also correlated with the source flux $f_s$ and with
the blended light $f_b$ that does not participate in the event, via
the equation for flux evolution 
\begin{equation}
F(t) = f_s A[u(t)] + f_b;\quad
u^2(t) = u_0^2 + {(t-t_0)^2\over t_\e^2},
\label{eqn:flux}
\end{equation}
and the magnification $A$ is given in Equation~(\ref{eqn:apm}).
Because $t_\obs$ is known exactly and both $t_0$ and $t_\eff$ are usually
extremely well-measured, $\phi$ can be determined very well.
Hence, the application of the measurements of both the magnitude
and the direction $\Delta\btheta_{-,+}$ to the interpretation of the
microlensing event depend only very weakly on the precision of the
measurement of the lightcurve's \citet{pac86} parameters.  This
feature makes interferometric imaging extremely robust.

Unfortunately, a single epoch of interferometric imaging still leaves
an important ambiguity.  Because of the way that $u_0$ enters 
Equation~(\ref{eqn:flux}), it is generally the case that only its magnitude 
is measured, not its sign.  In practice, this means that, in the absence
of additional information, one does not know whether $\phi$ should
be ``added'' clockwise or counterclockwise to the direction of 
$\Delta\btheta_{-,+}$ to obtain
the direction $-\bmu_\rel$ (i.e., the source-lens relative proper motion).
In fact, this ambiguity can be resolved
simply by measuring $\Delta\btheta_{-,+}$ at a second epoch.  That is,
at late times $\bu\rightarrow -(t-t_0)\bmu_\rel/\theta_\e$.  Hence,
$\phi$ should be ``added'' to $\Delta\btheta_{-,+}$ with the same chirality
as the ``motion'' of $\Delta\btheta_{-,+}$ from the first to the second
epoch.

In the great majority of cases, it is impossible to tell at the time
of the microlensing event whether the lens is dark or luminous simply
because the lens is superposed on a relatively bright source.  In most
cases this determination can be made one or several decades after the event,
when the lens and source have separated sufficiently to separately resolve
them in high resolution imaging (e.g., \citealt{ob05169bat,ob05169ben}).
This delay will decrease by a factor of 3--5 with the advent of next-generation
(``30 meter'') telescopes, but it will still be several years.  After that
wait, the luminous lenses can be further studied but the dark lenses cannot.
Therefore, the study of dark isolated lenses requires aggressive observations
during the microlensing events, before it is known which lenses are dark.

An interesting class of microlenses whose nature remains to be determined
are those giving rise to ``domestic microlensing events'', i.e.,
events with sources within 1--2 kpc of the Sun and lying toward directions
other than the $\sim 100\,{\rm deg^2}$ that are intensively monitored
toward the Galactic bulge.  The optical depth for microlensing of these
nearby sources is order $\tau\sim 10^{-8}$, i.e., about 100 times lower
than toward the Galactic bulge.  Moreover, there are only $N\sim$ few million
stars in these regions with $V\la 14$, i.e., within the range of amateur
observers, who are the only ones who have monitored such large portions of
the sky until very recently.  Hence, even if these amateurs were 100\%
efficient, one would expect these events to be detected at only at a rate
$\Gamma \sim (2/\pi)N\tau/(t_\e A_\min)\sim 0.5\,{\rm yr}^{-1}/A_\min$,
where $A_\min$ is the minimum magnification for detection.  These
statistics were noted by \citet{domestic}
when the first such $A\approx40$ event was found by an amateur astronomer A, Tago \citep{fukui07}.
Of course, with only one such event, nothing definite could be concluded.

Recently, a second such event, TCP J05074264+2447555 (RA=$05^{\rm h}07^{\rm m}42.\!\!^{\rm{s}}72$, Dec=$+24^\circ47'56.\!\!''4$, \citealt{gaiadr2}; hereafter referred to as 
``TCP J0507+2447'' for brevity), with magnification $A\approx10$
was found.  At this point, the detection of these events still does not strongly
contradict theoretical expectations, but reconciliation of theory and 
experiment does require a somewhat greater efficiency of amateur observers
than one might naively expect.  Thus, this potential contradiction led
our team (as well as many other astronomers, see, e.g., \citealt{nucita18}) to undertake very aggressive
observations in order to constrain the nature of this rare event.  In our case,
we undertook observations with the GRAVITY instrument \citep{grav17} of the Very Large Telescope Interferometer (VLTI) to obtain the first measurement of $\theta_\e$ by interferometric resolution of microlensed images. 

\section{{Observations and Data Reduction}
\label{sec:obs}}

In this section, we discuss the observations and data reduction of TCP J0507+2447 with a focus on the observing strategy and data reduction of VLTI GRAVITY. {{The readers who are not familiar with optical interferometry and its terminologies may refer to \citet{Lawson00} for detailed discussions and \citet{vltimicrolens2} for an introduction to interferometry observables in the context of microlensing.}}

\subsection{Observations}
\label{sec:obs_strat}

\begin{figure*}
\plotone{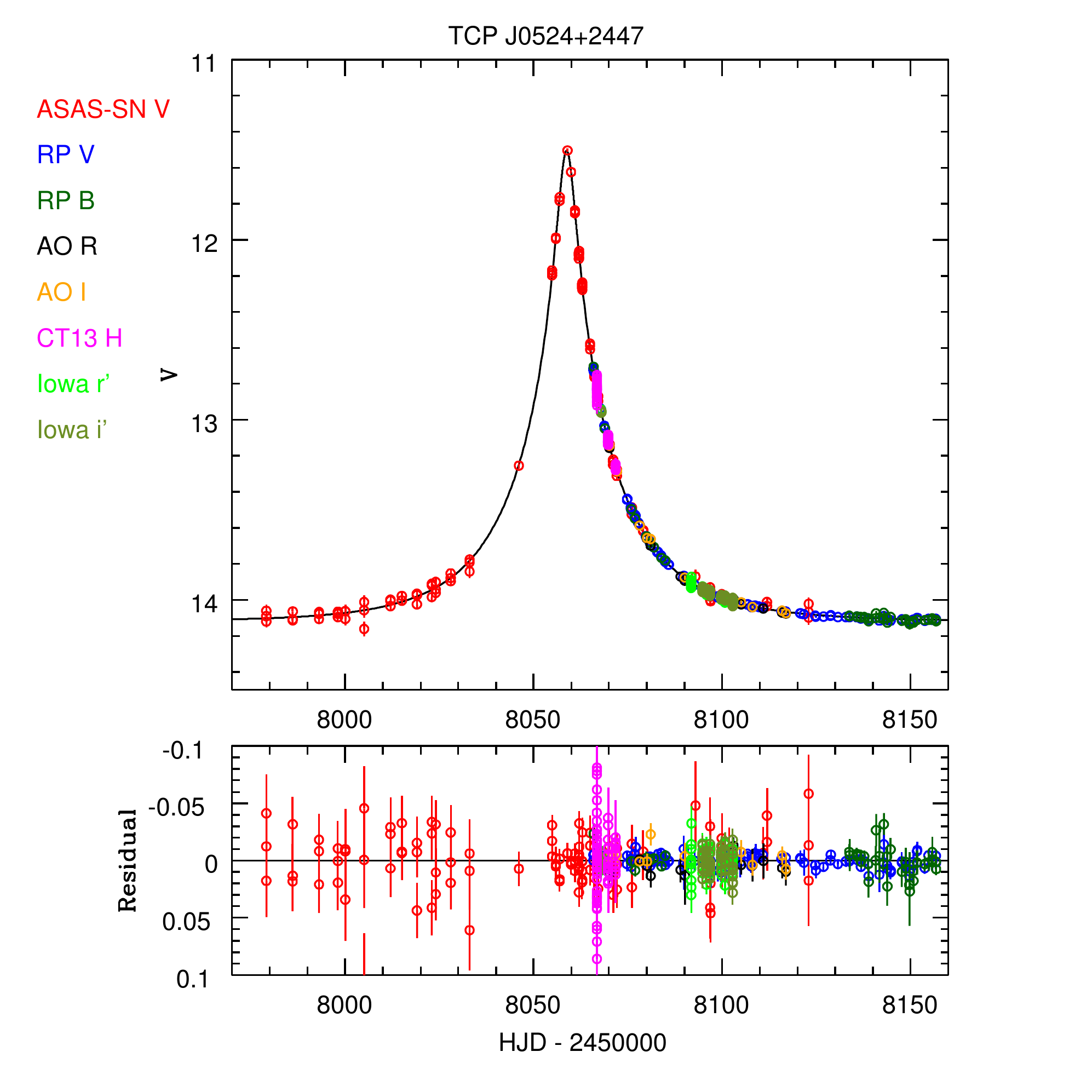}
\caption{Multi-band light curves of TCP J0507+2447 and the best-fit single-lens model. The data are from ASAS-SN $V$, Post Observatory (RP) in $V$ and $B$, Auckland Observatory (AO) $R$ and $I$, the 1.3m SMARTS telescope at CTIO (CT13) $H$ and the 0.5~m Iowa Robotic Telescope (Iowa) in bands very close to Sloan $r'$ and $i'$. All data are photometrically aligned with the $V$-band data using the microlensing model.}
\label{fig:lc}
\end{figure*}

The brightening of TCP J0507+2447 was first discovered by the Japanese amateur astronomer 
Tadashi Kojim (Gunma-ken, Japan) on UT 2017-10-25.688, and the discovery was reported to 
CBAT ``Transient Object Followup Reports''\footnote{\url{http://www.cbat.eps.harvard.edu/unconf/followups/J05074264+2447555.html}} on UT 2017-10-31.734.
The microlensing nature of the event was recognized using the data from All-Sky 
Automatic Survey for Supernovae (ASAS-SN; \citealt{asassn}) made available via ASAS-SN Light Curve Server v1.0 \citep{csk17}. \citet{ulensatel} found that the ASAS-SN $V$-band light curve ending on UT 2017-11-02.41 was consistent with a single-lens microlensing model. Subsequent multi-band imaging follow-up observations were performed at numerous sites. In this work, we use the follow-up data taken with 0.6~m telescopes at Post Observatory (CA and NM, USA) operated by R. Post (RP) in Johnson $BV$, the 0.5~m Iowa Robotic Telescope (Iowa) at the Winer Observatory (AZ, USA) in AstroDon E-series Tru-balance R and Andover 650FH90-50 longpass filters (which are very similar to Sloan $r'i'$ filters and will be referred to as $r'i'$ throughout the text),  the 0.4m telescope of Auckland Observatory (AO) at Auckland (New Zealand) in $RI$, and the 1.3m SMARTS telescope at CTIO (CT13) in $H$. The $H$-band data are calibrated with the Two Micron All Sky Survey (2MASS; \citealt{2mass}), and we add the archival 2MASS $H$-band measurement of $11.845\pm0.022$ at the baseline to the $H$-band light curve when modeling the light curve. For all optical and near-infrared (NIR) data, aperture photometry is performed following standard procedures. The light curves are shown in Figure~\ref{fig:lc}. The analysis including more follow-up data, in particular observations with {\it Spitzer} to obtain space-based microlens parallax constraints \citep{refsdal66,gould94a,dong07}, will be reported in a subsequent paper (Zang et al. in prep).

We examined the feasibility of observing this event with the VLTI: the peak limiting magnitude of $K\sim10$ was observable with 
the GRAVITY instrument \citep{grav17} on the VLTI using the 8-meter Unit Telescopes (UTs). 
The observation would be at the limit of feasibility due to the low observable altitude of the target ($< 40\deg$) and thus its high airmass. Furthermore, the position of the object on the sky
also implied that the projection of the baselines on the sky was very much contracted in one 
direction, giving rise to a lowered spatial resolution in that direction. Despite these limits we 
decided to submit a Director Discretionary Time (DDT) proposal to the European 
Southern Observatory. The DDT proposal was accepted on November 6 as DDT 2100.C-5014. 
The observation was planned for the first possible observation slot available on the UTs, 
on the nights of November 7 and 8, 2017.

The observation was performed by the operational Paranal team led by K. Tristram and X. Hautbois. On the first night, the seeing was excellent. Using the standard mode of GRAVITY 
(single-feed, visible adaptive optics and fringe tracking on the same star), the fringes could 
be found easily, but the fringe tracking loop could not be closed continuously. Therefore the 
observation was a partial success. The data can be used with some limitations explained below. On the second night, the target was fainter and the seeing was worse. The fringes could be briefly seen but could not be tracked, so no useful data were obtained during that night.

\subsection{{VLTI GRAVITY Data Reduction}
\label{sec:datared}}

We reduced the GRAVITY data using the standard data reduction software (DRS) v1.0.5 \citep{lap14}. The fringe-tracker data were reduced successfully. However, because 
the observations conditions were far from ideal, we obtained 
a locking ratio of only $\sim$80\%, whereas for typical good quality data, this value would
be above 99\%. The science spectrograph was running with an exposure time of 30 seconds, as a result no frames were recorded with continuous fringe tracking. As a result, all frames from the science spectrograph were rejected because the data reduction software is tuned for the high signal-to-noise (SNR) regime where the fringe tracker is continuously locked. The fringe tracker yields reduced complex visibilities, i.e., amplitude and phase. In principle, both the amplitude and the phase of the visibilities could provide useful information.
However, because the data were taken at the end of the night,
there were no post-observation calibration observations. This fact,
combined with the low locking ratio just mentioned, together implied
that the visibility amplitudes could not be used.  Hence, our
analysis rests entirely on the four 3-telescope closure phases. 

The calibrator observed with the microlensing event was {BD+15 788}, a $K=8.1$\,mag K0III type star, which we selected using the  \texttt{searchCal}\footnote{\url{http://www.jmmc.fr/searchcal\_page.htm}} \citep{cheli16} tool from the Jean-Marie Mariotti Center. The photometric estimate of the angular diameter is 0.12 mas, which means that the object is unresolved for the purpose of our observations. Based on the 90\% percentile injected flux in the fringe tracker we estimate that, the overall flux from the microlensing event was 10 times fainter, hence leading to $K\approx10.6$\,mag at the time of observation, 

Inspecting the closure phase data shows that the bluest spectral channel is very noisy, which leads to two very different values between our two observation data sets (which are 10 minutes apart). This is also true for the reddest channel for telescope triangles U4U3U2 and U3U2U1. We subsequently rejected these data points in our analysis. Concerning the error bars given by the DRS, the calibration errors have to be taken into account manually so we added quadratically 0.5 degrees of error, which corresponds to the scatter between our two calibrator data sets.  

Ideally, for a four-telescopes configuration, only three closure phases out of four are independent: the fourth being a linear combination of the three others. This is the case if the signal chain is affected only by phase biases per beam (i.e., we ignore closure phase instrumental biases) and if the reduction chain is the symmetrical for all closures. The closure phase redundancy can be broken if the 2-telescopes phases involved in two different closures calculation are not equivalent: for instance, they can be averaged differently when the two closures are averaged. In the case of our data, the combination of four closures has residuals of 1.4 degrees on average, with a scatter of 1.6 degrees RMS. This means that our four closures are not perfectly redundant. We hence chose to fit the four sets of closures, instead of a set of three, since each set of three would result in a different set of parameters.

\section{Constraints from the Photometric Light Curves}
\label{sec:anal}

The photometric light curves of TCP J0507+2447 and the best-fit single-lens model \footnote{\citet{nucita18} reported a short-lived planetary anomaly  that they modeled as a companion with planet-to-star mass ratio $q = 1.1\pm0.1\times10^{-4}$. Our data do not confirm or rule out the planetary deviations from the single-lens model because we did not have the necessary light-curve coverage over the reported anomaly. Nevertheless, we have conducted numerical experiments using the best-fit planetary microlensing models reported in \citet{nucita18} and found that the single-lens microlensing parameters relevant to our analysis show negligible changes if we add the reported planetary companion. We also find that the expected magnification and the positions of the major and minor images at the time of the VLTI observations are not affected, and this is because the VLTI data were taken more than 1 week after the reported planetary signals, which occurred at ${\rm HJD} \sim 2458058.5$.} are shown in Figure~\ref{fig:lc}. 
In addition
to  the 5 parameters $(t_0,u_0,t_\e,f_s,f_b)$ of the \citet{pac86} model, we incorporate the microlens parallax effects, parameterized by $(\pi_{\rm E,E}, \pi_{\rm E,N})$ \citep{gould04} in the fits. We also tried a model with zero blending for all but the ASAS-SN data (ASAS-SN's resolution is $16\arcsec$ and blending from ambient stars cannot be ignored). Blended light is detected at high statistical significance with $\Delta{\chi^2} = 182.5$ for 8 additional parameters. The results for both free blending and zero blending models are reported in Table~\ref{tab:single}. 

\begin{table}
\begin{center}
\begin{tabular}{ccccc}
\hline
\hline
Parameters        & Free blending & Zero blending \\
\hline
$t_0$ (HJD)-2450000 & $8058.76\pm0.01$ & $8058.74\pm{0.01}$\\
$u_0$                  & $0.084\pm0.001$ & $0.089\pm{0.001}$\\
$t_{\rm E}$ (days)& $27.92\pm{0.38}$ & $26.46\pm{0.08}$\\
$\pi_{\rm E, E}$  & $0.05\pm{0.15}$ & $0.56\pm{0.07}$\\
$\pi_{\rm E, N}$  & $0.20\pm{0.84}$ & $0.19\pm{0.42}$\\
$f_{b,B}/f_{s,B}$ & $ 0.043 \pm 0.017$ & 0\\
$f_{b,V}/f_{s,V}$ & $0.063 \pm 0.017$ & 0\\
$f_{b,r'}/f_{s,r'}$  & $0.083 \pm 0.017$& 0\\
$f_{b,R}/f_{s,R}$ & $0.105 \pm 0.018$& 0\\
$f_{b,i'}/f_{s,i'}$   & $0.112 \pm 0.018$& 0\\
$f_{b,I}/f_{s,I}$    & $0.141\pm 0.019$& 0\\
$f_{b,H}/f_{s,H}$ & $0.182 \pm 0.018$& 0\\
\hline
$A_{\rm VLTI}$   & $3.87\pm{0.05}$ & $3.687\pm{0.004}$\\
$u_{\rm VLTI}$   & $0.266\pm{0.003}$ & $0.2791\pm{0.0003}$\\
$\chi^2$ &  542.6  & 725.1\\
\hline 
\end{tabular}
\end{center}
\caption{Best-fit parameters and uncertainties for single-lens models with free blending parameters and fixed zero blending parameters. The predicted magnifications at the time of VLTI observations (${\rm HJD}_{\rm VLTI} = 2458065.8$) $A_{\rm VLTI}$ and $u_{\rm VLTI}$ are also reported. In total, there are 539 data points, and the best-fit $\chi^2$ values for both model are reported too.}
\label{tab:single}
\end{table}

The VLTI measurements are made in $K$ with a field of view (FOV) of $\sim 0\arcsec.05$. The measured blending fraction $f_b/f_s$ from the light curves increases with wavelength. A linear relation between $f_b/f_s$ and the logarithm of the filter central wavelength fits the trend well, and we estimate $f_{b,K}/f_{s,K} = 0.20\pm0.02$ for the VLTI bandpass. This implies that the blend is $K \approx 13.6$ based on the 2MASS baseline $K=11.680\pm0.018$. In principle, this blended light could be due to an ambient star within the $\sim 1 \arcsec$ point spread function (PSF) of the non-ASAS-SN photometric follow-up observations. However, the 2MASS surface density of sources with $K<13.7$ toward this field is only $\approx 3\times10^{-4}$ arcsec$^{-2}$, and the prior probability of finding an unassociated star within a $1 \arcsec$ PSF is only $\sim 10^{-3}$. Therefore, the blend is likely associated with the 
microlensing event: either it is the lens itself, a companion to the lens, a companion to the source or some combination of these possibilities. If the blend were 
due to the lens, it would lie inside the VLTI FOV. If the blend were a luminous companion to either the lens or the source and within the VLTI FOV, they would face severe restrictions from the microlensing light curves in the form of binary-lens perturbation or binary-source signals. The companion might instead be distant enough to be outside the VLTI FOV and unconstrained from the light curve modeling. We defer the detailed discussion on the light-curve constraints on the companion to a follow-up paper (Zang et al. in prep).  In the following section, we model the VLTI data by assuming both possibilities. The magnification at the time of VLTI observations is well constrained to be $A_{\rm VLTI} = 3.87\pm0.05$ (i.e., $u_{\rm VLTI} = 0.266\pm0.003$). 

\begin{figure*}
\plotone{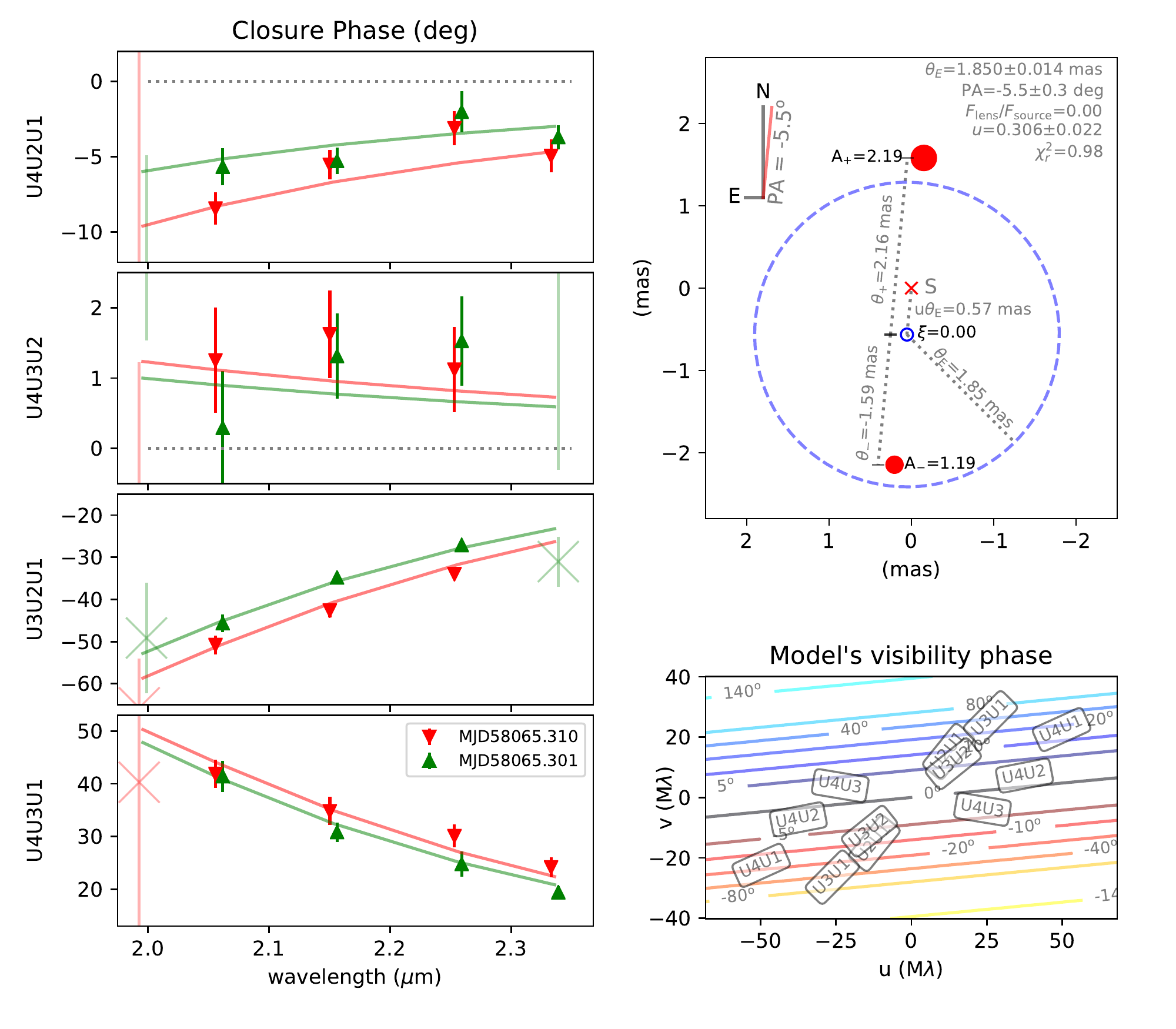}
\caption{VLTI model with no lens light. Left: closure phase data (points and error bars) and model (continuous lines) for the four sets of telescope triangles, as function of wavelength. The 2 data sets, represented by 2 different symbols, have been slightly offset in wavelength for clarity. Upper Right: model of the apparent images. The two red dots are the major and minor images (note that the size of the dots does not represent the actual apparent sizes of the images), the 'x' symbol in red is the un-lensed source position (labeled "S"). The lens position is shown as a blue open circle, and the dashed circle is the Einstein ring. The flux of each component (minor/major images, lens) is given in fractions of un-lensed source flux. Lower Right: visibility and phase function for the best model in the $u,v$ plane, as well as our 4T configurations at the time of observations. The colored lines are contour lines showing the visibility phase (in degrees), blue for positive phase and red for negative phase, indicating the orientation of the "binary".}
\label{fig:model_zero}
\end{figure*}

\section{{Modeling the Interferometric Data}
\label{sec:model}}

Because this is a microlensing event, there are guaranteed to be
at least three effectively point sources in the field, i.e., the
lens and the two images of the source. As discussed above, the blend detected from 
light-curve modeling can be due to the lens or luminous companions to the lens and/or source
outside the VLTI FOV.  We have limited ability to directly address these with the 
interferometric data.  Hence, we proceed to analyze the VLTI data using two sets of models: 1) assuming that 
there is negligible light from the lens (``no lens light'') and 2) assuming that the blend is the lens (``luminous lens'').

\subsection{{Simplest Model: No Lens Light}
\label{sec:nolens}}

In our first analysis, we assume that there is no light from the lens,
so that the FOV of VLTI contains only a ``binary star'', formed by the two lensed images. We treated these images as unresolved by the interferometer. We apply
the well-verified \texttt{CANDID} software\footnote{\url{https://github.com/amerand/CANDID}} \citep{gall15} 
to model the closure phases and find
\begin{equation}
\begin{aligned}
&\Delta\theta_{-,+}  = 3.75\pm 0.03\,\mas,
\qquad
\psi = -5.5^\circ \pm 0.3^\circ; \\\qquad
&\eta = 0.546\pm 0.032,
\end{aligned}
\label{eqn:nolensparms}
\end{equation}
where $\Delta\theta_{-,+} $ is the scalar separation, $\psi$ is the position
angle (North through East), and $\eta$ is the flux ratio. The $\chi^2$ of the fit is 25.5. Since we have two data sets, each with 14 data points, and three fitted parameters, the number of degrees of freedom is 25, yielding a reduced $\chi^2$ of 0.98. Figure~\ref{fig:model_zero} shows the closure phase data and the best-fit model (left), and the 4-telescope (4T) configuration at the time of observations. The visibility and phase function for the best model in the $u, v$ plane are  shown in the lower right panel of Figure~\ref{fig:model_zero}.

Without referencing the microlensing light curves, we can estimate the magnification,
$A=(1+\eta)/(1-\eta) = 3.38\pm 0.21$. Inverting Equation~(\ref{eqn:apm}), this yields
\begin{equation} 
u=\sqrt{2/\sqrt{1-A^{-2}}-2}=0.306\pm 0.022
\end{equation}

Hence,
\begin{equation} 
\theta_\e = {1\over\sqrt{1 + u^2/4}}{\Delta\theta_{-,+} \over 2}= 1.850\pm 0.014\,\mas
\label{eqn:nolensthetae}
\end{equation}
Note in particular that the fractional uncertainty in the first factor in 
Equation~(\ref{eqn:nolensthetae}) is only 0.16\%.  Hence, the error in
the estimate of $\theta_\e$ is completely dominated by the error in
$\Delta\theta_{-,+} $.  The model of the apparent image is illustrated in the 
upper right panel of Figure~\ref{fig:model_zero}.

The derived $u=0.305\pm 0.020 $ is in $2\sigma$ tension with the value estimated from light-curve modeling $u_{\rm VLTI} = 0.266\pm0.003$. We also attempt to impose the prior $u = 0.266\pm0.003$ to the \texttt{CANDID} models. The best-fit model has a only marginal increase of $\Delta{\chi^2} = 1.75$, and the best-fit values $\theta_\e = 1.837\pm 0.012\,\mas$ and $\psi = -5.14^\circ \pm 0.29^\circ$ are within $1\,\sigma$ of those given in Equations~(\ref{eqn:nolensparms}) and (\ref{eqn:nolensthetae}). Therefore, the VLTI data have much weaker 
constraint on $u$/magnification, and imposing the magnification prior from the photometric models hardly change the 
derived $\theta_\e$ and $\psi$.

\texttt{CANDID} allows free searches for 
``third body companions'' to binary stars.  We conducted such a search
but did not find any statistically significant local minima.  Of course, this does
not rule out the presence of other sources: it only means that we cannot
probe for other sources without specifying their positions.  The other
potential source for which we actually know the position is the lens.
So we next analyze the data within the context of such 
``restricted 3-body'' models.

\subsection{{Models With a Luminous Lens}
\label{sec:lenslight}}

To conduct such modeling, we re-parameterize the problem so that the 
lens will automatically be positioned correctly relative to the two
images of the source.  We use parameters $(\theta_\e,\psi,u,\xi)$,
where $\xi=f_l/f_s$ and $f_l$ is the flux from the lens and $f_s$ 
is the flux of the source, un-amplified.  Then,
from Equation~(\ref{eqn:apm}), we infer $A_\pm$, and from 
Equation~(\ref{eqn:upm}), we infer $\theta_\pm$, i.e., separations
from the lens (which is at origin) along opposite directions specified
by $\psi$.  The ratio of the ``secondary'' (minor-image) to the ``primary'' (major-image) flux
is then $\eta=(A-1)/(A+1)$, while the ratio of lens flux to the ``primary'' 
is $2\xi/(A+1)$. 
\begin{figure*}
\includegraphics[scale = 0.7]{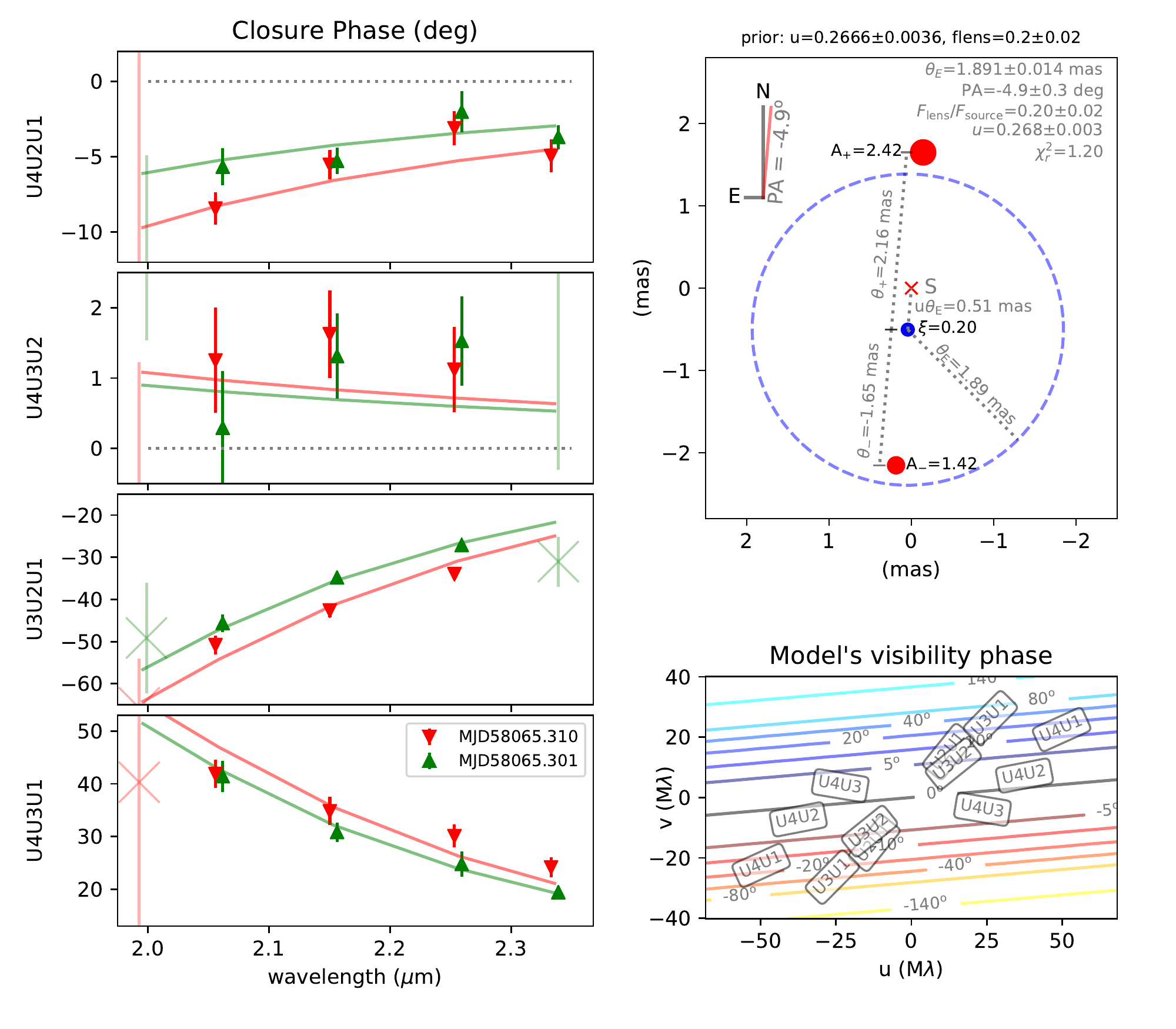}
\caption{The VLTI model with a luminous lens. The panels are the same as in Figure~\ref{fig:model_zero}.}
\label{fig:model_ll}
\end{figure*}

We impose the priors that $u = u_{VLTI} = 0.266\pm0.003$ and $\xi = f_{b,K}/f_{s,K} = 0.20\pm0.02$ based on the 
best-fit photometric models. The best-fit model, as illustrated in Figure~\ref{fig:model_ll}, has a reduced $\chi^2$ of 1.20, 
which is $\Delta{\chi^2} = 5.5$ worse than the ``no lens light'' VLTI model with no external prior.  The best-fit parameters  are,
\begin{equation}
\theta_\e = 1.891\pm 0.014\,\mas;\,\, \psi = -4.9^\circ \pm 0.3^\circ.
\label{eq:three}
\end{equation}

These values are only $\approx2\,\sigma$ different from the values in Equations~(\ref{eqn:nolensthetae}) and~(\ref{eqn:nolensparms}) of the ``no lens light'' VLTI model. 

Therefore, the present VLTI data only weakly constrain the magnification and lens flux. The interferometric 
angular Einstein radius is robustly estimated to the $1-2\%$ level, independent of whether or not the lens contributes to the flux detected in the VLTI FOV.

Finally, we determine $\phi$, the angle between the instantaneous source-lens separation, $u$, and the direction of source-lens relative proper motion, $-\mathbf{\mu}_\mathrm{rel}$. Recall that this quantity depends only on the photometric light curves, via the parameters $t_0$ and $t_\mathrm{eff}$ . We find ($t_0$, $t_\mathrm{eff}$) = (8058.76, 2.35) $\pm$ (0.01, 0.02), and these yield,
\begin{equation}
\phi = \cot^{-1}\frac{\delta t}{t_\mathrm{eff}} = 18.4^{\circ}\pm 0.2^{\circ}.
\end{equation}

\begin{figure}
\plotone{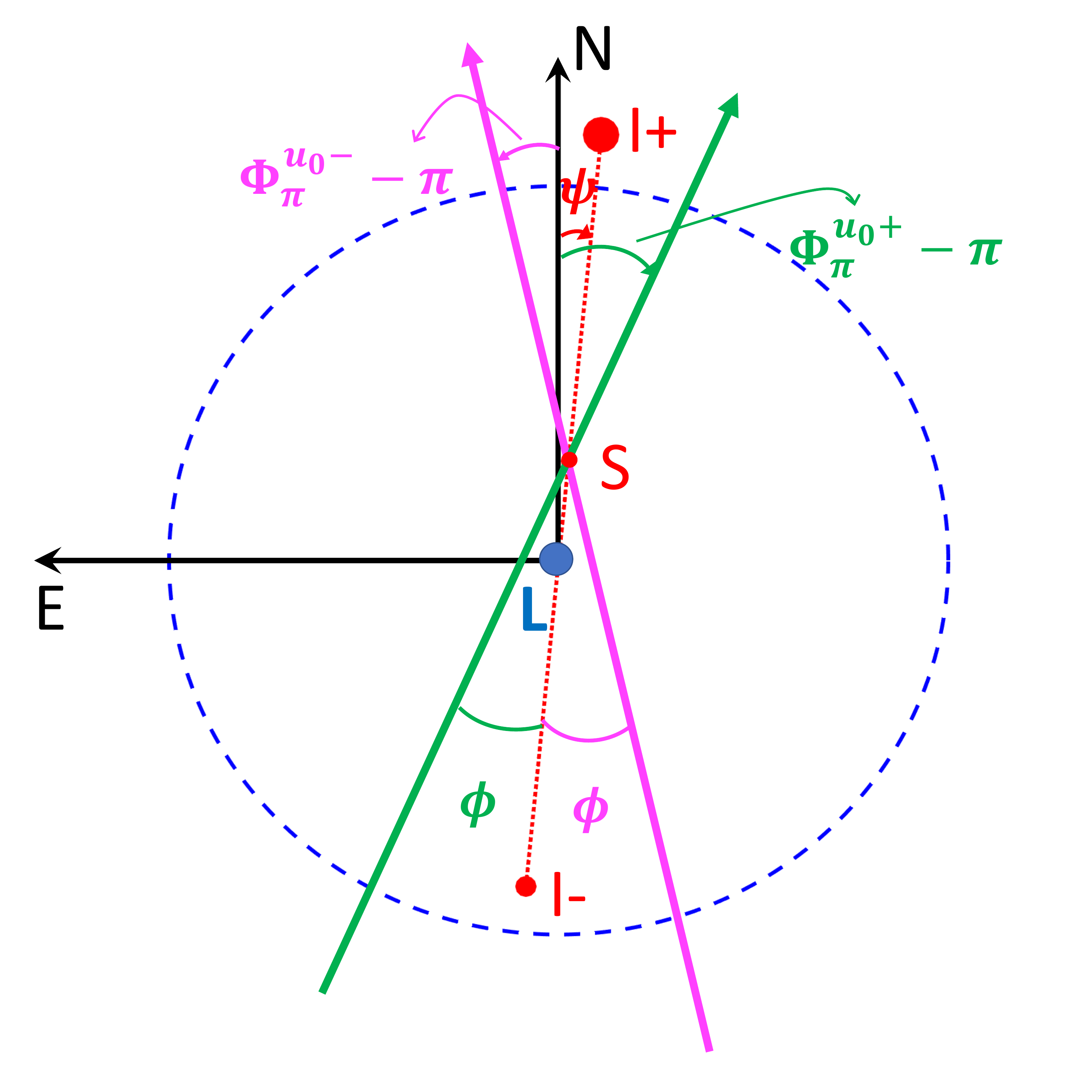}
\caption{The geometry of the microlensing model. North is up and East is to the left. The lens (``L'') position is shown as a blue dot, and the Einstein ring is represented by a circle in blue dashed line. The source position at the VLTI measurement is shown as a red dot labelled with ``S'', and the two images are shown as two red dots labelled with ``I$+$'' (major) and ``I$-$'' (minor). The position angle of the two images from North through East is defined as $\psi$, and this is a VLTI observable. There are two degenerate solutions of the lens-source trajectory angle  $\Phi_\pi$ measured from North through East, and the two trajectories are shown in blue and magenta lines, respectively, with the arrows indicating the direction of the source-lens relative proper motion. The angle $\phi$ between the source-lens relative proper motion and
source-lens relative position can be determined directly from the
light-curve parameters (see Eq.~\ref{eqn:phiu}). The two degenerate solutions are $\Phi_\pi^{u_0+} = \psi-\phi+\pi$ and $\Phi_\pi^{u_0-} = \psi+\phi+\pi$ for positive and negative $u_0$, respectively. Note that, in this plot, an angle with a clockwise arrow has a negative value, and the angle $\phi$ is always positive by definition.
}
\label{fig:geometry}
\end{figure}

As discussed in Section~\ref{sec:intro}, because we have interferometric data at only a single epoch, we cannot determine whether $\phi$ should be added or subtracted from $\psi$ to find the direction of relative source-lens motion. That is, this direction could be either $12.9^{\circ}\pm 0.4^{\circ}$ or $-23.9^{\circ} \pm 0.4^{\circ}$ (North through East) based on the ``no lens light''  model; and either $13.5^{\circ}\pm 0.4^{\circ}$ or $-23.3^{\circ} \pm 0.4^{\circ}$ (North through East) for the ``luminous lens'' model. As discussed in Section~\ref{sec:intro}, the direction of the microlens parallax $\bpi_\e$ is defined as that of lens-source proper motion, and thus its direction $\Phi_\pi^{u_0+} = \pi + \theta - \psi$ (for the positive $u_0$ solution) or $\Phi_\pi^{u_0-} = \pi + \theta + \psi$ (for the negative $u_0$ solution). The relevant geometry is shown in Fig.~\ref{fig:geometry}. We discuss in Section~\ref{sec:mass} how this ambiguity can ultimately be resolved using \textit{Gaia} and/or \textit{Spitzer} data.

\section{Discussion}
\label{sec:discussion}

{In summary, using VLTI GRAVITY, we successfully resolve the images of a microlensing event 
for the first time. As a result, we obtain a precise measurement of the angular Einstein 
radius regardless of whether or not a luminous lens contributes to the flux detected in the VLTI FOV: $\theta_\e = 1.850\pm 0.014\,\mas$ for the ``no lens light'' model and $\theta_\e = 1.883\pm 0.014\,\mas$ for the ``luminous lens'' model.

\subsection{Comparison with the \citet{nucita18} $\theta_\e$ measurement}
\label{sec:nu}

\citet{nucita18} reported a short-duration planetary anomaly near the peak of their TCP J0507+2447 light curves, and by modeling the finite-source effects when the source passed by the planetary caustics, they measured $\rho\equiv\theta_*/\theta_{\rm E} = (6.0\pm 0.8)\times 10^{-3}$, where $\theta_*$ is the angular size of the source. \citet{nucita18} made a rough estimate of $\theta_*$ by assuming that the source distance is in the range 
of $700$-$800\,{\rm pc}$. In what follows, we use the tight empirical color/surface-brightness relation of \citet{kervella04} to estimate $\theta_*$, which is a commonly adopted approach in microlensing works \citep{ob03262}.

To estimate the source radius $\theta_*$, we first evaluate the apparent magnitude of the source in two bands, $V_s$ and $K_s$. For the former, we simply adopt the value from the fit to the microlensing event, $V_s=14.207$.  For the latter, we first note the 2MASS baseline value $K_{\rm base}=11.68$ and then take account of the $f_{b,K}/f_{s,K}=0.20$ blending in $K$-band to derive $K_s=11.88$.  Hence, the observed color is $(V-K)_s = 2.327$.  Next we note that the CMB extinction toward this direction is $A_{V,\rm CMB}=1.579$ and $A_{K,\rm CMB}=0.174$ \citep{extinction},
implying $E(V-K)_{\rm CMB}=1.405$.  The source has been spectroscopically typed as F5V \citep{spectype}, for which $(V-K)_{s,0}=1.08$ \citep{mscolor}.  We infer that the source lies behind a fraction $\lambda=[(V-K)_s-(V-K)_0]/E(V-K)_{\rm cmb}=0.89$ of the CMB extinction. The source (l=178.756, b=-9.325) is about 125\,pc above the plane by using $\pi_{\rm Gaia}=1.45\,$mas, and given that the dust scale height is about 100 pc and also that the dust density declines toward the direction of the source (i.e., in the anti-center direction), this value of $\lambda$ is in the expected range. Hence, we find that the de-reddened source fluxes are $V_{s,0}=12.80$ and $K_{s,0}=11.72$.  Using the color/surface-brightness relation of Kervella et al. (2004),
we obtain $\theta_* = 9.0\,\mu$as, and the overall error on this procedure is about 10\%.  We note that if we simply use the values $R_{*,\rm Gaia}=1.49\,R_\odot$ and $\pi_{\rm Gaia}=1.45\,$mas based on Gaia DR 2, then we 
obtain $\theta_* = 10.0\,\mu$as.  However, given that the Gaia ``star'' is actually a blend of the source and a closer and redder blend, this result is consistent with the estimate based on the color/surface-brightness relation.

We then combine this result with the measurement by \citet{nucita18} of $\rho\equiv\theta_*/\theta_{\rm E} = (6.0\pm 0.8)\times 10^{-3}$, to  obtain $\theta_{\rm E} = 1.50\pm 0.25\,$mas.  This compares to the much more
precisely derived value of $\theta_{\rm E}=1.87\pm  0.03\,$mas (considering both the ``no lens light'' and ``luminous lens" solutions) 
in the present work. Therefore, we find that the two agree at the $1.5\,\sigma$ level. This agreement gives added confidence to the planetary interpretation of the anomaly found by \citet{nucita18}.}

\subsection{Future mass, distance, and transverse velocity measurements}
\label{sec:mass}

Recall from Equation~(\ref{eqn:mpirelmurel}) that by measuring $\theta_\e$
and $\bpi_\e$, one can immediately determine $(M,\pi_\rel,\bmu_\rel)$.
Since the source star is bright, the {\it Gaia} second data release (DR2)
already has a very accurate measurement of $\pi_s$ and $\bmu_s$: $\pi_{Gaia} = 1.451\pm0.031\,{\rm mas}$ and $(\mu_{\rm RA, {Gaia}} = -0.229\pm0.061\,{\rm mas/yr}, \mu_{\rm Dec, {Gaia}} = -7.33\pm0.033\,{\rm mas/yr})$.\footnote{More precisely, what Gaia actually measures is the flux-weighted mean
parallax and proper motion of all sources within the Gaia
point spread function (PSF), ${\rm FWHM}_{Gaia} \sim 100\,$mas.
Because $\theta_\e \ll {\rm FWHM}_{Gaia}$, this implies that
the Gaia parallaxes and proper motions will always be weighted
means of the source and lens flux:
$\pi_{Gaia} = (\pi_s + r_{Gaia}\pi_l)/(1 + r_{Gaia})$,
where $r_{Gaia}=f_{l,Gaia}/f_{s,Gaia}$, and similarly for $\bmu_{Gaia}$.
Note that if (as in the present case) $r_{Gaia}\ll 1$, then
$\pi_{Gaia} \simeq \pi_s + r_{Gaia}\pi_\rel$.  If we now identify
$f_l=f_b$, then using Table 1, we obtain
$r_{Gaia}\simeq (r_V + r_B)/2 \simeq 0.053$.  This implies that
the fractional correction to the naive Gaia parallax is
$\simeq 5.3\%(\pi_\rel/\pi_s)$, which is likely to be of order or
somewhat larger than the Gaia statistical error.   Therefore,
this correction must be taken into account once $\pi_\rel=\pi_\e\theta_\e$
is measured.}
Hence, this would immediately yield the lens distance $D_l=\au/(\pi_\rel+\pi_s)$
and transverse velocity $\bv_\perp = D_L(\bmu_\rel+\bmu_s)$.  Because
GRAVITY has already measured $\theta_\e$, the only missing ingredient is 
$\bpi_\e$.

As briefly mentioned in Section~\ref{sec:intro}, it is generally
substantially easier to measure $\bpi_\e$ if its
direction $\Phi_\pi$ is known independently.  As we discuss below, this
will be case for TCP J0507+2447.  As also mentioned in Section~\ref{sec:intro}, 
two epochs of interferometric imaging yield this direction, but a single
epoch determines the direction only up to a two-fold ambiguity,
$\Phi_\pi = \pi + (\psi \pm \phi)$.  As illustrated in Fig.~\ref{fig:geometry}, 
here $\psi$ is the position angle
of the major image with respect to the minor image and $\phi$ is the
angle between the source-lens separation vector and the direction
of source-lens relative motion.

Unfortunately, as noted in Section~\ref{sec:obs}, interferometric
data could only be obtained over an interval of a few minutes.
Hence this ambiguity in $\Phi_\pi$ remains.

\subsubsection{{Resolution of the Direction Ambiguity with {\it Gaia}}
\label{sec:gaia}}

To understand how {\it Gaia} can resolve the directional ambiguity,
we begin by considering a series of $N$ astrometric measurements carried
out uniformly over time $T$, each with error $\sigma_0$ in each direction,
and with the conditions $T/N\ll 9 t_\e \ll T$.  (We explain the reason
for the ``9'' further below).\ \ 
Because $N\gg 1$,
the parallax and proper motion of the source will be measured with
much higher precision than the individual measurements, and indeed because
$9 t_\e\ll T$, they will be measured much better than any astrometric
quantity that can be derived from measurements during ${\cal O}(9 t_\e)$.
Hence, for our purposes, we can regard the true position of the source
as ``known perfectly''.  The error in the measurement of the offset
of the image centroid from the source is then simply the error in the former.
Hence, we can write the SNR of the ensemble of 
measurements of these offsets as
\begin{equation}
\begin{aligned}
({\rm SNR})^2 &= \sum_{i=1}^N {u_i^2 \theta_\e^2\over (u_i^2 + 2)^2\sigma_0^2}\\
&= \sum_{i=1}^N {(u_0^2 +(t_i-t_0)^2/t_\e^2)\theta_\e^2\over 
(u_0^2 +(t_i-t_0)^2/t_\e^2 + 2)^2\sigma_0^2} .
\end{aligned}
\label{eqn:snr}
\end{equation}
In most cases of interest (including the present one), $u_0^2\ll 2$.
Hence we can approximate $u_0\rightarrow 0$. Making use $T/N \ll 9 t_\e$,
we can approximate the sum as an integral, and then making use
of $9 t_\e\ll T$ we can take the limits of this integral to infinity,
\begin{equation}
\begin{aligned}
({\rm SNR})^2 &\rightarrow 
{\theta_\e^2 N\over \sigma_0^2}{t_\e\over T}
\int_{(-T/2-t_0)/t_\e}^{(T/2-t_0)/t_\e} d\tau{\tau^2\over (\tau^2+2)^2} \\
&\simeq {\theta_\e^2 N\over \sigma_0^2}{t_\e\over T}
\int_{-\infty}^\infty d\tau{\tau^2\over (\tau^2+2)^2},
\label{eqn:snr2}
\end{aligned}
\end{equation}
where $\tau\equiv (t-t_0)/t_\e$.   This is easily evaluated
\begin{equation}
{\rm SNR} \rightarrow {\theta_\e\over\sigma_{\rm tot}}
\sqrt{{t_\e\over T}\,{\pi\over 8^{1/2}}}
\simeq {\theta_\e\over\sigma_{\rm tot}}
\sqrt{t_\e\over T}
\label{eqn:snr3}
\end{equation}
where $\sigma_{\rm tot}\equiv \sigma_0/\sqrt{N}$ is the astrometric
precision of the position measurement from the entire series
of observations.  From Equation~(\ref{eqn:snr2}) we see that the
maximum value of the integrand is 1/8, while from
Equation~(\ref{eqn:snr3}) we see that the value of the integral
is $(\pi/8^{1/2})t_\e$.  Hence the effective width of the integral
is the ratio of these, i.e., $t_{\rm width}= 8^{1/2}\pi t_\e\simeq 9 t_\e$.
This is the reason that the quantity defining the extreme limits is
``$9 t_\e$'' rather than simply ``$t_\e$''.

The fractional error in the $\theta_\e$ measurement as well as
the angular error (in radians) of $\Phi_\pi$ are both equal to (SNR)$^{-1}$.

For {\it Gaia} measurements of TCP J0507+2447, the condition $u_0^2\ll 2$ is
well satisfied.  To evaluate the conditions,
$T/N\ll 9 t_\e\ll T$, we first note that $9 t_\e\sim 0.7\,$yr, compared to
$T=5\,$yr for the baseline {\it Gaia} mission.  Hence, $9 t_\e\ll T$ is 
reasonably satisfied.  Second, for targets near the ecliptic,
{\it Gaia} can be expected to make about 35 visits
(each composed of two observations) over 5 years, which are restricted
to $\sim 70\%$ of the year by the Sun exclusion angle.  Hence
$(T/N)_\eff\sim 0.1\,$yr.  Hence, $T/N\ll 9 t_\eff$ is also reasonably
satisfied.

If we adopt a 5-year mission precision of {\it Gaia} for a $V=14$
source of $\sigma_{\rm tot}\sim 25\,\muas$, then we obtain
\begin{equation}
\sigma(\Phi_\pi) \simeq {25\,\muas\over 1.8\,\mas}
\sqrt{27\,{\rm day}\over 5\,{\rm yr}}{\,57^\circ\over {\rm rad}}
= 6.6^\circ .
\label{eqn:sigmaphi}
\end{equation}
This error bar should be compared to the difference in the angles
between the two solutions $2\phi=37^\circ$.  Hence, it is very likely
that {\it Gaia} can resolve the ambiguity in $\Phi_\pi$.  Recall
that because of the much higher precision of the GRAVITY measurement
of $\psi$, the role of {\it Gaia} is only to break the ambiguity,
not actually measure $\Phi_\pi$.

For completeness, we note the following facts about uniform astrometric
monitoring of microlensing events, even though they are not directly
relevant to the case of TCP J0507+2447.  First, if one were to take account of
improved astrometry of magnified images (due to the fact that they
are brighter), this would lead to a modification of Equation~(\ref{eqn:snr3}).
If (as in the case of TCP J0507+2447), the astrometric errors are decreased
by a factor $A^{-1/2}$, then one easily finds that the continuous
formula changes by $\pi/2^{3/2}\rightarrow 2^{1/2}\ln(2^{1/2}+1)$.  On the
other hand, for much fainter (``below sky'') targets for which
the errors are reduced by $A^{-1}$, we have
$\pi/2^{3/2}\rightarrow \pi/2$.  The main importance of these formulae
is that they are not very different, i.e., they lead to improvements
of SNR by factors 1.12 and 1.19, respectively.  However, the main
reason for not using the first formulae here is that its conditions
for use are [$(T/N \ll t_\e)$ \& $(9 t_\e\ll T)$] (rather than
$T/N \ll 9 t_\e\ll T$).  This condition is not met by {\it Gaia} for
TCP J0507+2447 and indeed will never be met for any {\it Gaia} event because
it implicitly requires $N\gg 9^2$.  However, the second (``below sky'')
formula only requires $T/N \ll t_\e\ll T$ and so might plausibly
be met by some faint, {\it Gaia}-microlensing black-hole candidates.

Finally, we note that Equation~(\ref{eqn:snr3}) can easily be generalized
to the case of $u_0\not= 0$ by 
$\pi/2^{3/2}\rightarrow \pi(1+u_0^2)/(2+u_0^2)^{3/2}$.  It is of some
interest to note that this formula peaks at $u_0=1$ and that it
only falls back to $\pi/2^{3/2}$ at $u_0^2=(1+\sqrt{5})/2$.  This
means that astrometric events can be detected and well measured
even when there is no obvious photometric event.  The ``below sky''
formula is also easily generalized: $\pi/2\rightarrow \pi/\sqrt{4+u_0^2}$.
Unfortunately, the generalization of the ``above sky'' formula can
not be written in closed form.

\subsubsection{{Future Parallax Measurement}
\label{sec:parallax}}

Unfortunately, the photometric light curves do not yield useful
parallax information.  The event is quite short compared to a year,
so only if $\pi_\e$ were extremely large would we expect a full
measurement of $\bpi_\e$ from the annual-parallax effect.  Nevertheless,
one might have hoped to measure the component parallel to
Earth's acceleration, $\pi_{\e,\parallel}$, which would induce
an asymmetry in the light curve.

However, the event lies quite close to the ecliptic and it peaked
only three weeks from opposition.  Hence, the component of Earth's acceleration
transverse to the line of sight is only 1/3 of its full amplitude.  In addition,
by chance the lens-source relative proper motion points roughly south
whereas Earth's acceleration points roughly east.  Combined, these factors
imply that the light-curve asymmetry induced by Earth's acceleration
is only about 1/10 of what it could be for the most favorable geometry.
Thus, it is not
surprising that there is no detectable signal.

Fortunately, {\it Spitzer} has taken a series of observations covering
its visibility window, $1.7< (t-t_{0,\oplus})/t_\e < 3.1$.  As we now
argue, it will probably not be possible to properly interpret these
observations until $\Phi_\pi$ is determined by combining {\it Gaia}
data with our interferometric measurement (as described in 
Section~\ref{sec:gaia}).  Hence, we present here the general principles
of such a measurement, which would be the first from such late-time
{\it Spitzer} observations, and we explain how these rest critically on the
precision of the measurement that we have made of $\psi$.

In general, space-based parallax measurements
derive from a time series of space-based
photometric measurements.  If these measurements cover the peak and
wings of the event as seen from space, then one can directly measure
$t_{0,\rm sat}$ and $u_{0,\rm sat}$ from the light curve.  Then by comparing
these to $t_{0,\oplus}$ and $u_{0,\oplus}$  measured from the ground
(and knowing the projected separation of the satellite from Earth,
${\bf D}_\perp$), one can determine $\bpi_\e$ \citep{refsdal66,gould94a}.
If (as is very often the case for {\it Spitzer} observations), only
the post-peak tail of the light curve is observed from space, then
it is impossible to measure $t_{0,\rm sat}$ and $u_{0,\rm sat}$ from the
satellite light curve alone.  However, using color-color relations
linking the ground-based and space-based data, one can determine
the space-based source flux independent of the space-based light curve.
With this added information, it is possible to extract the full $\bpi_\e$
from even a post-peak light curve \citep{170event}.

However, the case of TCP J0507+2447 is substantially more extreme than
any previous one.  We know that at the first observation 
$\tau\equiv(t-t_{0,\oplus})/t_\e=1.7$.  Moreover, although we do not
know the precise angle of $\Phi_\pi = \pi+(\psi\pm\phi)$, we do know
that it is roughly due south, whereas {\it Spitzer} lies roughly
to the west.  Hence, the Einstein-ring distances $\tau$ and
$\pi_\e D_\perp/\au$ add approximately in quadrature.  So, for
example, if $\pi_\e=1$, then (with $D_\perp\sim 1.4\,\au$),
$u_{\rm sat}\sim 2.2$ at the first observation.  This would imply
$A=1.045$.  While such a small magnification is likely measurable
for the bright source in this event, it is less clear that $(t_0,u_0)$ could
be reconstructed from a falling light curve starting at such a low
level.

Fortunately, this is not actually necessary.  Because we know (or,
after the {\it Gaia} measurement, will know) $\Phi_\pi$ very 
well, and we know the direction of {\it Spitzer} extremely well, we will
also know the angle between them.  This will then permit us to employ a
variant of the idea proposed by \citet{gould12} (and verified by
\citealt{ob161045}) of ``cheap space-based parallaxes''.   They showed
that with just one or two space-based photometric measurements
taken near $t_{0,\oplus}$ (plus a late-time ``baseline'' measurement)
one could measure $\pi_\e$ for high-magnification $(u_0\ll 1)$ events.

We now show that the same is true even if the measurements are taken
well after peak, {\it provided} that one knows $\Phi_\pi$ independently.
Let $\gamma$ be the (known) angle between the satellite and $\bpi_\e$.
Then, from the law of cosines,
\begin{equation}
\begin{aligned}
u_{\rm sat}^2 = & (\tau_\oplus-u_{0,\oplus}\cos\gamma)^2 + 
\biggl({D_\perp \pi_\e\over\au}+u_{0,\oplus}\sin\gamma\biggr)^2 \\
&-2(\tau_\oplus-u_{0,\oplus}\cos\gamma)
\biggl({D_\perp \pi_\e\over\au}+u_{0,\oplus}\sin\gamma\biggr)\cos\gamma .
\end{aligned}
\label{eqn:cosines}
\end{equation}
By measuring $A$ at some time $A(t) = (F(t)-F_{\rm base})/F_s$,
where the $F$ are flux measurements and $F_s$ is the independently
determined source flux, and inverting Equation~(\ref{eqn:apm}), one
can determine $u_{\rm sat}^2$ and so (by inverting
Equation~(\ref{eqn:cosines})), determine $\pi_\e$. 

This sequence of calculations has a number of potential ambiguities,
which we now discuss.  First, from the ground-based light curve,
only the magnitude of $u_{0,\oplus}$ is known, but not its sign.
Hence, inversion of Equation~(\ref{eqn:cosines}) will yield two
different answers depending on the sign chosen for $u_0$.  However,
if $\tau$ and/or $D_\perp \pi_\e/\au$ is large compared to $u_0$,
then the impact on the measurement of $\pi_\e$ will be small.  This
is why the method is restricted to ``high-magnification events'',
similarly to \citet{gould12}.

Applying this condition by setting $u_{0,\oplus}$ to zero, we can
invert Equation~(\ref{eqn:cosines}) to obtain
\begin{equation}
\pi_\e ={\au\over D_\perp}\biggl(\tau_\oplus\cos\gamma \pm 
\sqrt{u^2_\sat - \tau^2\sin^2\gamma}\biggr)
\label{eqn:cosinvert}
\end{equation}
 From Equation~(\ref{eqn:cosinvert}), one can immediately see
that if $\gamma$ is acute, then there are two positive solutions,
whereas if it is obtuse, then there is only one.  (Note that $\pi_\e$ is
positive definite because the direction of $\bpi_\e$ is known.)\ \
This also implies that if $\pi_\e\sim (\au/D_\perp)\tau_\oplus\cos\gamma$,
then the error in the estimate of $\pi_\e$ will be large.  However,
both of these problems can be countered if there are additional data
taken further on the decline.

At the present time, there are two values for $\gamma$ that differ by
$2\phi\simeq 37^\circ$.  This immediately implies that 
Equation~(\ref{eqn:cosinvert}) cannot be uniquely interpreted
until this ambiguity is broken by {\it Gaia}.  In the approximation
that $\sin\gamma$ is the same for both of these solutions, the difference
in $\pi_\e$ is roughly $\Delta\pi_\e\sim 0.9$, which is quite large.
In principle, it
is possible that the ambiguity could be resolved by the time
series of {\it Spitzer} measurements, but this may prove difficult.

In brief, the very precise measurement of $\Phi_\pi$ that can be achieved
either by two epochs of interferometric imaging or one such epoch 
combined with a {\it Gaia} astrometric microlensing measurement,
can enable satellite parallax measurements, even under much less favorable
conditions than has heretofore been possible.

\acknowledgments 
We thank the operational Paranal team led by K. Tristram and X. Hautbois. We thank Akihiko Fukui for making us aware of T. Kojima's discovery and Szymon Kozlowski for his help.

S.D. and P.C. acknowledge Projects 11573003 supported by the National Science Foundation of China (NSFC). This work was partly supported by NSFC 11721303. This research has made use of the Jean-Marie Mariotti Center \texttt{SearchCal} service \footnote{Available at http://www.jmmc.fr/searchcal} co-developped by LAGRANGE and IPAG, and of CDS Astronomical Databases SIMBAD and VIZIER \footnote{Available at http://cdsweb.u-strasbg.fr/}. This research has also made use of the Jean-Marie Mariotti Center \texttt{Aspro2} service \citep{2013ascl.soft10005B}\footnote{Available at http://www.jmmc.fr/aspro2}. Work by AG was supported by AST-1516842 from the US NSF and by JPL grant 1500811. AG received support from the European Research Council under the
European Union's Seventh Framework Programme (FP 7) ERC Grant Agreement n. [321035]. 

We thank the Las Cumbres Observatory and its staff for its continuing support of the ASAS-SN project. We are grateful to M. Hardesty of the OSU ASC technology group. ASAS-SN is supported by the Gordon and Betty Moore Foundation through grant GBMF5490 to the Ohio State University and NSF grant AST-1515927. Development of ASAS-SN has  been supported by NSF grant AST-0908816, the Mt. Cuba Astronomical Foundation, the Center for Cosmology and AstroParticle Physics at the Ohio State University, the Chinese Academy of Sciences South America Center for Astronomy (CAS- SACA), the Villum Foundation, and George Skestos. 

This paper uses data products produced by the OIR Telescope Data Center, supported by the Smithsonian Astrophysical Observatory. This work has made use of data from the European Space Agency (ESA) mission
{\it Gaia} (\url{https://www.cosmos.esa.int/gaia}), processed by the {\it Gaia}
Data Processing and Analysis Consortium (DPAC,
\url{https://www.cosmos.esa.int/web/gaia/dpac/consortium}). Funding for the DPAC
has been provided by national institutions, in particular the institutions
participating in the {\it Gaia} Multilateral Agreement.

\software{DRS (v1.0.5; \citealt{lap14}), 
CANDID \citep{gall15}}

\end{document}